\documentclass[twocolumn,aps,prl,superscriptaddress]{revtex4}

\usepackage{amsmath}
\usepackage{graphicx}
\usepackage{MnSymbol}

\begin{document}

\title{On the validity of the method of reduction of dimensionality: area of contact, average interfacial separation and contact stiffness}

\author{I.A. Lyashenko}
\affiliation{Peter Gr\"unberg Institut-1, FZ-J\"ulich, 52425 J\"ulich, Germany}
\affiliation{Sumy State University, Rimskii-Korsakov Str. 2, 40007 Sumy, Ukraine}
\author{Lars Pastewka}
\affiliation{Dept. of Physics and Astronomy, Johns Hopkins University, Baltimore, MD 21218, USA}
\affiliation{Fraunhofer-Institut f\"ur Werkstoffmechanik IWM, Freiburg, 79108 Germany}
\author{Bo N. J. Persson}
\affiliation{Peter Gr\"unberg Institut-1, FZ-J\"ulich, 52425 J\"ulich, Germany}

\begin{abstract}
It has recently been suggested that many contact mechanics problems between 
solids can be accurately studied by mapping the problem on an effective 
one dimensional (1D) elastic foundation model. Using this 1D mapping
we calculate the contact area and the average interfacial separation between elastic 
solids with nominally flat but randomly rough surfaces. We show,
by comparison to exact numerical results, that the 1D mapping method
fails even qualitatively. We also calculate the normal interfacial 
stiffness $K$ and compare it with the result of an analytical study.
We attribute the failure of the elastic foundation model to the neglect of the
long-range elastic coupling between the asperity contact regions. 
\end{abstract}

\maketitle

\pagestyle{empty}


{\bf 1 Introduction}

\begin{figure}
\includegraphics[width=0.8\columnwidth]{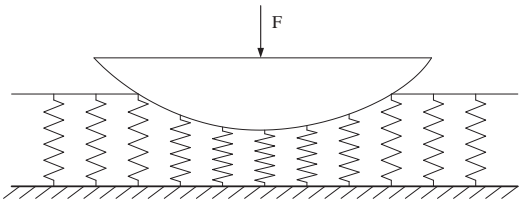}
\caption{\label{Drawing}
In the elastic foundation model the elastic solid is replaced by an array of independent springs.
}
\end{figure}

The calculation of the stress and displacement field resulting from the contact 
between elastic solids with rough surfaces is a very complex problem, in part
due to the many length scales usually involved, and also because of the long-range
elastic coupling between the contact regions. For this reason simplifying
approaches are very important. However, most analytical theories, such as the
Greenwood-Williamson contact mechanics theory\cite{Greenwood66}, or theories based on the elastic 
foundation model\cite{foundation} (see Fig. \ref{Drawing}), neglect the elastic coupling
between asperity contact regions. It has recently been shown by 
exact numerical studies that the neglect of the elastic coupling results in
qualitatively wrong contact topography\cite{Almqvist11}, and even the relation between the 
contact force and the area of contact is incorrectly described using this approach\cite{Carbone08}.
In Ref. \cite{Ramisetti11,Campana08} it was also shown that the contact stress-stress
correlation function (in wavevector space) $\langle \sigma({\bf q}) \sigma(-{\bf q})\rangle \sim q^{-\alpha }$,
where in the overlap model $\alpha=2+H$ (where $H$ is the Hurst exponent), while including
the long-range elastic coupling $\alpha=1+H$.

In a series of papers, Popov and coworkers have proposed that a simple 1D-elastic foundation
model can be used to accurately describe the contact between elastic solids\cite{Geike,Pohrt12,Pohrt12PRE}.
In a recent publication they calculated the normal stiffness between elastic solids with
randomly rough but nominally flat surfaces, and argued that the results are in good agreement with exact numerical results\cite{Pohrt12PRE}. 
In this note we will show that in fact the model of
Popov et al fails even qualitatively to describe the contact mechanics correctly.

\vskip 0.3cm
{\bf 2 Area of real contact and the average interfacial separation}

Consider two elastic solids with nominally flat surfaces squeezed together 
by the nominal pressure $p=F/A_0$. The average interfacial separation is denoted by $\bar u$.
As $p$ increases, the average interfacial separation $\bar u$ monotonically decreases, while the area
of real contact $A$ increases~\cite{persson01,hyun04,Berthoud98,Barber03}.
In earlier publications~\cite{Persson07,hyun04,Campana07,Robbins11,Campana10} it has been shown
that in a large pressure range $A\sim p$ and $\bar u \sim {\rm ln} p$.
This can be understood as follows: 
As the load increases, existing contact patches grow and new, small contacts are formed.
This happens in such a way that the distribution of contact sizes and
local pressures remains approximately constant
over a wide range of loads\cite{hyun04,Campana07}.
It follows that $A\sim p$ and that the elastic deformation energy (per unit nominal contact area), $U_{\rm el}$, 
stored at the interface must be proportional to the load or the nominal contact pressure\cite{Persson07}:
\begin{equation}
\label{eq:elastEnerg}
U_{\rm el} = u_0 p ,
\end{equation}
where $u_0$ is a length parameter of order the root-mean-square roughness $h_{\rm rms}$.
Since the elastic energy is equal to the work done by the external load
(assuming hard-wall interactions and no adhesion), it follows that
$$p=-{dU_{\rm el}\over d\bar u}.$$
Combining this with (1) gives
\begin{equation}
p = p_0 {\rm exp} (-\bar u/u_0), \label{eq:oldEqno1}
\end{equation}
where $p_0$ is an integration constant.
The theory of Persson predicts that $u_0=\alpha h_{\rm rms}$ and $p_0 = \beta E^*$, where $E^*$ is the
effective elastic modulus and $\alpha$ (of order unity) and $\beta$ are dimensionless.
Both $\alpha$ and $\beta$ only depend on the spectral properties
of the surface\cite{Persson07,Robbins11,Almqvist11,Carbone11,Carbone09,Campana10}.

In the same pressure range where (2) is valid, the area of real contact
\begin{equation}
{A\over A_0} = {\kappa \over \xi} {p\over E^*}, \label{Area}
\end{equation}
where $\xi = \langle (\nabla h)^2 \rangle^{1/2}$ is the surface rms-slope and $\kappa \approx 2$. 

Eqs. (2) and (3) are only valid at such high pressures that multi-asperity contact occurs. 
At very low pressures the solids will only make contact in the vicinity of the highest
asperity. In this finite-size pressure region the relation between $\bar u $ and $p$ will exhibit large
fluctuations from one surface realization to another\cite{Lorenz}. 
In the study presented below the finite-size region is too small to be observed on the linear pressure scale
used in Fig. \ref{new.all}. In Ref. \cite{PRLS,PRL} we have studied numerically and analytically the relation between
the interfacial stiffness and the squeezing pressure in both the finite size pressure region and for higher pressures,
and in Sec. 5 we compared the results for the stiffness with the 1D-elastic foundation model of Popov et al.

\begin{figure}[thbp]
\includegraphics[width=0.7\columnwidth]{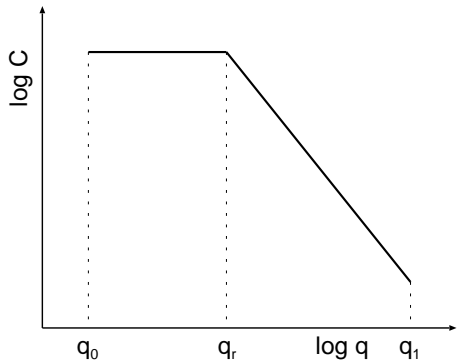}
\caption{\label{roll}
The surface roughness power spectrum as a function of the wavevector (log-log scale) for a self-affine fractal
surface with a roll-off.
}
\end{figure}

\vskip 0.3cm
{\bf 3 Numerical results for $A(p)$ and $\bar u (p)$}

In the reduction of dimensionality approach a 3D-contact problem is mapped on a 1D-elastic 
foundation problem. Here we are interested in the contact between
two nominally flat but randomly rough surfaces. For frictionless contact, this problem can be mapped
on an elastic half space with a randomly rough surface in contact with a rigid substrate with a flat surface. 
In the contact mechanics theory of Popov et al. the roughness of the 
1D-substrate has a power spectrum related to that of the original via the equation:
\begin{equation}
C_{\rm 1D} (q) = \pi q C_{\rm 2D}(q).
\label{Cq}
\end{equation}
The spring constant of the elastic foundation
is related to the effective (or combined) elastic modulus via $k=aE^*$, where $a$ is the spacing
between the springs. Using standard procedures we have generated randomly
rough 1D-surfaces with the power spectra given by (4). As in an earlier study\cite{PRL}, the original
2D surface is self affine fractal with the Hurst exponent $H=0.7$ (or fractal dimension $D_{\rm f} = 3-H = 2.3$)
and with the small and large cut-off wavevectors $q_0 = 1$ and $q_1 = 8192$. We consider two cases, namely
when the substrate surface is fractal-like in the whole interval $q_0 < q <q_1$, and when there is a roll-off
at $q_{\rm r} =8$, see Fig. \ref{roll}.

The red and blue solid lines in Fig. 3 have been calculated following the procedure outlined by Popov et al.\cite{pop,pop1,pbook,PhMes}:
Each independent spring is compressed into compliance by the 1D rough surface profile where the profile overlaps with the initial relaxed spring positions.
In each step we calculated the force $F$ and 
area of contact $A$. The applied force $F$ was calculated as the sum of forces for all springs in contact:
\begin{equation}
F = k\sum\limits_{i=1}^n \Delta x_i,
\label{Force}
\end{equation}
where $n$ is the number of springs in contact, $\Delta x_i$ is the spring compression. 
After this we have calculated the area of contact $A$~\cite{pop,pbook,PhMes}: 
\begin{equation}
A = \frac{\pi}{4} \sum\limits_{i=1}^{n_{\rm c}}(an_i)^2, 
\end{equation}
where $n_{\rm c}$ is the number of connected regions. In this case all springs in connected regions must be in contact, $n_i$ is the number of 
springs in each region, $an_i$ are the diameters of these 
regions. 
In the case of full contact the area of contact $A=A_0$, where
\begin{equation}
A_0 = \frac{\pi}{4}(aN)^2,
\label{A_zero} 
\end{equation}
where $N$ is the full number of springs. Then there is only a single connected region, 
the diameter of which is equal to the length of the system $aN$.  
Using \eqref{Force} and \eqref{A_zero} we can also calculate the squeezing pressure $p=F/A_0$. 
The interfacial separation $\bar u$ was calculated using the formula: 
\begin{equation}
\bar u = \frac{1}{N}\sum\limits_{i=1}^{N}u_i,
\end{equation}
where $u_i$ is the distance between the end of each spring and the substrate surface. For springs in contact $u_i = 0$. 

Using this simple procedure we have calculated the dependencies shown in Fig. 3 by the solid lines. This dependencies were calculated 
in range of coordinates of elastic foundation (array of springs) from first contact to full contact. 

\begin{figure}
\includegraphics[width=0.9\columnwidth]{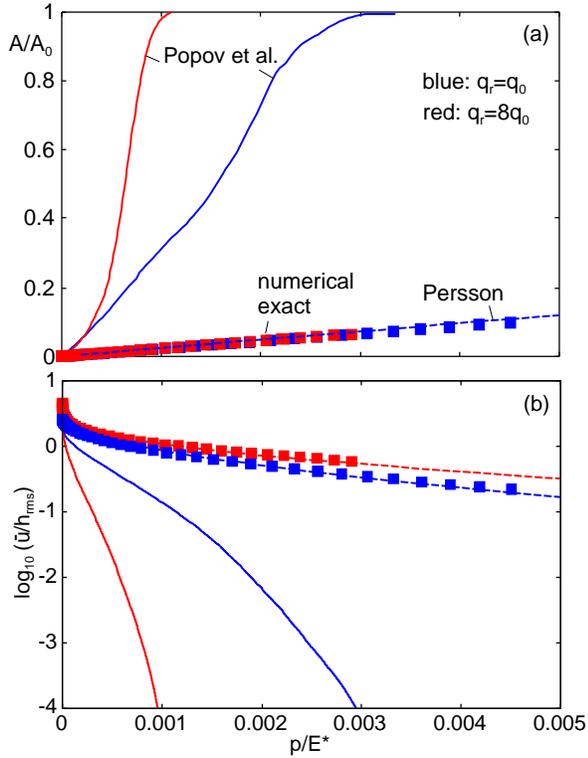}
\caption{\label{new.all1}
(Color online)
(a) The area of real contact $A$ in units of the nominal contact area $A_0$, and (b) the average interfacial separation $\bar u$
in units of the rms roughness $h_{\rm rms}$, 
as a function of the squeezing pressure $p$ in units of the effective elastic modulus $E^*$.
For self-affine fractal surfaces with $H=0.7$ and rms-slope 0.1. The surfaces have the small
and large wavevector cut-off $q_0=1$ and $q_1=8192$, respectively, and the roll-off
wavevector $q_{\rm r} =1$ (blue curves) and $q_{\rm r}=8$ (red curves). 
}
\end{figure}

We will compare the predictions of the theory of Popov et al. with numerical exact results obtained as described in Ref. \cite{exact1,exact2}.
In brief, this method computes the surface displacements using a Fourier-transform technique with a linear surface Green's function that corresponds to Poisson ratio $\nu=1/2$.
The interaction with the rigid surface is treated as a hard-wall repulsion.

In Fig. \ref{new.all1}(a) we show the calculated normalized contact area $A/A_0$ as a function of the squeezing pressure. 
The red and blue squares are the result of a numerical exact study and the dashed line the prediction using the theory of Persson.
Since the two surfaces have the same rms-slope the theory predict the same curve for both cases which agree well with the numerically exact results. 
The red and blue solid lines are the predictions of the theory of Popov et al.
Since $A(p)$ approach $A_0$ much faster in the model by Popov et al. then in the numerically exact theory, 
the interfacial stiffness $K=-dp/d\bar u$ will approach infinity much faster (with increasing
pressure) in the theory of Popov et al., as compared to our exact numerical study. Thus the stiffness relation $K(p)$ will also be incorrectly given by the 
theory of Popov et al. (see also Sec. 5). 

Fig. \ref{new.all1}(b) shows the logarithm of the average interfacial separation $\bar u$ as a function of the squeezing pressure $p$.
Again there is good agreement between the numerically exact results and the theory of Persson, while the approach of Popov et al. fail 
qualitatively. The results presented using the theory of Popov et al. are obtained by averaging the calculated quantities 
over 100 realizations of the rough-line topography, with the 1D-power spectrum given by (4). In each realization the elastic foundation has 8192 springs.

\vskip 0.3cm
{\bf 4 Contact stiffness}

Consider two elastic solids with nominally flat surfaces squeezed together 
by the nominal pressure $p=F/A_0$. 
From (\ref{eq:oldEqno1}) it follows that the contact stiffness 
$$K=-{d p \over d \bar u}= {p\over \alpha h_{\rm rms}}$$
or
$${K h_{\rm rms}\over E^*} = {1\over \alpha}{p\over E^*}\eqno(9)$$
This equation is only valid at such high pressures that multi-asperity contact occurs. 
At very low pressures the solids will only make contact in the vicinity of the highest
asperity. In this finite-size pressure region the relation between $K$ and $p$ will exhibit large
fluctuations from one surface realization to another\cite{Lorenz}.

In Ref. \cite{PRLS} two of us has derived an (approximate) analytical expression for
the (ensemble averaged) interfacial stiffness in the finite-size region.
The derivation assumes a self-affine fractal surface with the surface roughness power spectrum shown in Fig. \ref{roll}.
The surface is characterized by the Hurst exponent $H$ and the small and large wavevector cut-off $q_0$ and $q_1$, as well as a roll-off $q_{\rm r}$ (see Fig. \ref{roll}). 
For this model the stiffness per unit area in the low pressure, finite size, region is approximately given by (see Ref. \cite{PRLS})
$$K \approx \left ( {E^* \over L^2 q_{\rm r}} \right )^{H/(1+H)} \left ({p\over h_{\rm rms} }\right )^{1/(1+H)}\sim p^{1/(1+H)}$$
where $L \approx 2 \pi /q_0$ is the linear size of the studied system. Note that we can also write this equation as
$${K h_{\rm rms}\over E^*} \approx \left ( {h_{\rm rms} \over L^2 q_{\rm r}} \right )^{H/(1+H)} \left ({p\over E^*} \right )^{1/(1+H)}\eqno(10)$$

\vskip 0.3cm
{\bf 5 Numerical results for $K(p)$}

\begin{figure}
\includegraphics[width=0.9\columnwidth]{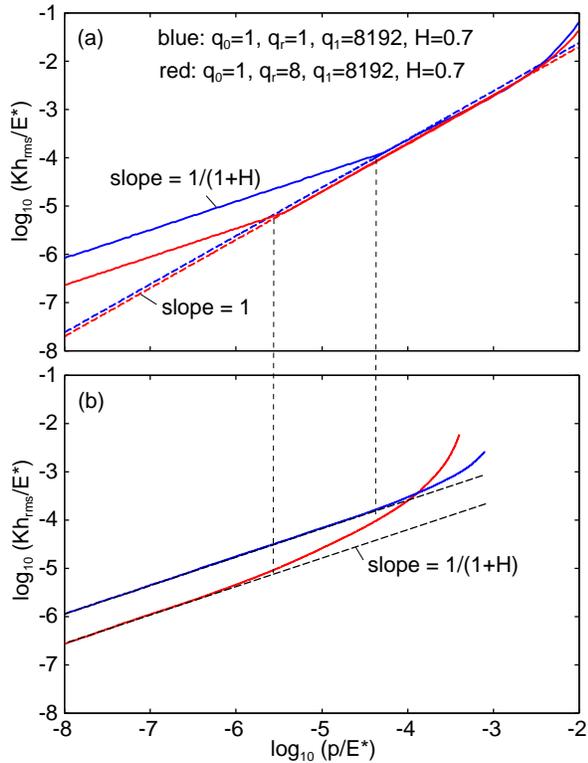}
\caption{\label{new.all}
(Color online)
Log-log plot of the nondimensional contact stiffness $Kh_{\rm rms}/E^*$  vs.
nondimensional pressure $p/E^*$ for self-affine
fractal surfaces with $H=0.7$ and rms slope 0.1. The surfaces have the small
and large wavevector cut-off $q_0=1$ and $q_1=8192$, respectively, and the roll-off
wavevector $q_{\rm r} =1$ (blue curves) and $q_{\rm r}=8$ (red curves). The result in (a)
is from the Persson contact mechanics theory, which agrees with the exact numerical
study presented in Ref. \cite{PRL}. The result in (b) is using the theory of Popov et al.
The vertical dashed lines indicate the pressures where the Popov et al. theory
starts to deviate from the analytical results.
}
\end{figure}

We consider again two cases,
when the substrate surface is fractal-like in the whole interval $q_0 < q <q_1$, and when there is a roll-off
at $q_{\rm r} =8$, see Fig. \ref{roll}.
In Fig. \ref{new.all}(a) we show the calculated interfacial stiffness using the theory of Persson. 
We have plotted $Kh_{\rm rms}/E^*$ as a function of $p/E^*$ as these dimensionless quantities enter in the theory
[see (9) and (10)]. The results in Fig.  \ref{new.all}(a) are in excellent agreement with exact numerical simulations for the same system, see
Fig. 1 in Ref. \cite{PRL}. One can distinguish three regions in the stiffness $K(p)$ relation. For very small
pressures the stiffness increases as $K\sim p^{1/(1+H)}$. This is a finite size effect, which occurs
when a single effective Hertz contact region, formed at the highest substrate asperity, prevails.
For higher pressures a region where $K\sim p$ is observed. 
This region, which becomes wider as the width of the roll-off region increases, 
results from contact with many asperities, and depends crucially on the long-range
elastic coupling between the contact regions. 
Finally, for very large pressure the interfacial separation approaches zero and the interfacial
stiffness increases towards infinite. 

Fig. \ref{new.all}(b) shows the results using the contact mechanics theory of Popov et al. 
The results are obtained by averaging the contact stiffness obtained in 100 realizations of the 
rough-line topography, with the 1D-power spectrum given by (4). In each realization the elastic foundation has 8192 springs.
Fig. \ref{new.all}(b) shows that the theory correctly predicts the initial (low pressure) relation $K\sim p^{1/(1+H)}$.
This result is expected because the study in Ref. \cite{PRLS} shows that $K\sim p^{1/(1+H)}$
holds even when one neglect the elastic coupling between the asperity contact regions.
However, the region where $K$ increases linear with the pressure $p$ is absent in Fig. \ref{new.all}(b). This is also 
expected because the $K\sim p$ result depends crucially on the elastic coupling between the asperity contact regions,
which is not included in the theory of Popov et al. 
As shown in Fig. \ref{new.all}(a), the linear region is particular large when there is a roll-off
in the power spectrum. Most surfaces of engineering interest exhibit a roll-off even larger
than for the $q_{\rm r}/q_0 = 8$ case shown in Fig. \ref{roll}. Thus in most practical applications, in particular involving
elastically soft materials like rubber, one will be in the linear $K\sim p$ region where 
the approach of Popov et al fails.

We note that wether there is a roll-off or a cut-off at $q_{\rm r}$ has very little influence on the result.
However, this does not imply that the only thing which matters
is the range over which the surface is self-affine fractal. The point is that including a roll-off
or cut-off at $q_{\rm r} > q_0$ impies roughly that the surface is ``periodically'' repeated $(q_{\rm r}/q_0)^2$ times.
This implies that there will be many asperities of similar height as the highest asperity.
This in turn means that the contact will much quicker (with increasing pressure) come into the multi-asperity contact
configuration where the stiffness $K$ depends linearly on the nominal squeezing pressure $p$.
This is the origin of why the linear relation between $K$ and $p$ starts at lower pressures, and extends over a larger pressure range,
when the surface has a roll-off or cut-off. 

The vertical dashed lines in Fig. \ref{new.all} indicate the pressures where the theory of Popov et al.
starts to deviate from the analytical results in Fig.  \ref{new.all}(a). Note that these points correspond to the start of the 
linear $K\sim p$ region in the analytical theory. This is expected as the linear region corresponds to multi-asperity contact,
where the elastic coupling between the asperity contact regions, which is neglected in the Popov et al. theory,  becomes important.
Thus it is the neglect of the long range elasticity, and not the reduction in dimensionality, which is the basic
problem with the approach of Popov et al. In fact, a recent study by Scaraggi et al.\cite{Scaraggi} has shown that if the
long-range elastic coupling is included in the analysis, it is possible to make 2D 
isotropic roughness approximately equivalent to 1D roughness.

\vskip 0.3cm
{\bf 6 Summary and conclusion}

We have presented a detailed comparison of the theory of Popov
et al. with numerical exact results and analytical results for self-affine fractal surfaces with and without
a roll-off. The theory of Popov et al. fails qualitatively to describe the $A(p)$ and $\bar u (p)$ relations, and 
we attribute this to the absence of the elastic coupling between the asperity contact regions in the approach of Popov et al.
In fact, a recent study by Scaraggi et al.\cite{Scaraggi} has shown that if the
long-range elastic coupling is included in the analysis, it is possible to make 2D
isotropic roughness approximately equivalent to 1D roughness.

We have also presented a detailed comparison of the theory of Popov
et al. with analytical results for the normal contact stiffness. For the case of a
roll-off at $q_{\rm r} =8$ 
the Persson theory and the exact numerical results presented in Ref. \cite{PRL}, exhibit
a linear $K\sim p$ region extending over 3
decades in pressure, while there is no linear region 
in the theory of Popov et al. The latter theory predicts $K\sim p^{1/(1+H)}$ in the limit of small pressures,
but this result is expected since 
an effective Hertz single-asperity contact prevails in this case (see Ref. \cite{PRLS}).
However, since the elastic coupling between the asperity contact regions is absent 
in the approach of Popov et al., no linear $K\sim p$ region is obtained.

\vskip 0.3cm
{\bf Acknowledgements}
We thank Giuseppe Carbone, Martin M\"user and Mark Robbins for useful discussions. 
L.P. acknowledges funding from the European Commission (Marie-Curie IOF-272619).

\end{document}